\documentclass[pra,twocolumn,floatfix,showpacs,showkeys,preprintnumbers,amssymb,amsmath,superscriptaddress]{revtex4}
\usepackage[latin1]{inputenc}
\usepackage{graphicx}% Include figure files
\usepackage{dcolumn}% Align table columns on decimal point
\usepackage{bm}% bold math
\usepackage[mathscr]{eucal}
\usepackage{epsfig}
\usepackage{rotating}
\usepackage[dvips]{color}

\begin{document}
\newcommand{\beq}{\begin{equation}}
\newcommand{\eeq}{\end{equation}}
\newcommand{\ket}{\rangle}
\newcommand{\bra}{\langle}
\newcommand{\A}{\mathbf{A}}
\preprint{ }
%\baselineskip{2.0}
\title{Implementing Shor's algorithm on Josephson Charge Qubits}

\author{Juha J. Vartiainen}
\email{juhav@focus.hut.fi}
\affiliation{Materials Physics Laboratory, POB 2200 (Technical
Physics),  Helsinki University of Technology, FIN-02015 HUT,
Finland}
\author{Antti\ O.\ Niskanen}
\affiliation{VTT Information Technology,
Microsensing, POB 1207, 02044 VTT, Finland}
\author{Mikio Nakahara}
\affiliation{Materials Physics Laboratory, POB 2200 (Technical
Physics), FIN-02015 HUT, Helsinki University of Technology,
Finland} \affiliation{Department of Physics, Kinki University,
Higashi-Osaka 577-8502, Japan}
\author{Martti\ M.\ Salomaa}
\affiliation{Materials Physics Laboratory, POB 2200 (Technical
Physics), FIN-02015 HUT, Helsinki University of Technology,
Finland}

\date{\today}% It is always

\begin{abstract}
We investigate the physical implementation of Shor's factorization algorithm on a
Josephson charge qubit register. While we pursue a universal method to factor
a composite integer of any size, the scheme is demonstrated for the number 21.
We consider both the physical and algorithmic requirements for an optimal implementation
when only a small number of qubits is available.
These aspects of quantum computation are usually the topics of
separate research communities; we present a unifying discussion of both of these fundamental
features bridging Shor's algorithm to its physical realization using Josephson junction qubits.
In order to meet the stringent requirements set by a short decoherence time, we accelerate the
algorithm by decomposing the quantum circuit into tailored two- and three-qubit gates
and we find their physical realizations through numerical optimization.
\end{abstract}



\maketitle

\section{Introduction}

%\subsection{Quantum computer}

Quantum computers have potentially superior computing power over
their classical counterparts~\cite{Galindo,Nielsen}. The novel computing
principles which are based on the quantum-mechanical superposition
of states and their entanglement manifest, for example, in Shor's
integer-factorization algorithm~\cite{Shor} and in Grover's
database search~\cite{Grover}. In this paper we focus on Shor's
algorithm which is important owing to its potential applications
in (de)cryptography. Many widely applied methods of public-key
cryptography are currently based on the RSA algorithm
\cite{Crypto} which relies on the computational difficulty of
factoring large integers.

Recently, remarkable progress towards the experimental realization of
a quantum computer has been accomplished, for instance, using nuclear spins
\cite{Steffen,Vandersypen}, trapped ions
\cite{Leibfried,Schmidt-Kaler}, cavity quantum electrodynamics
\cite{Yang}, electrons in quantum dots \cite{Wiel}, and
superconducting circuits
\cite{pc,Yu,Pashkin,nakamura,martinis,vion}.
However, the construction of a large multiqubit register remains extremely
challenging. The very many degrees of freedom
of the environment tend to become entangled with those of the
qubit register which results in undesirable decoherence~\cite{zurek}.
This imposes a limit on the coherent execution time
available for the quantum computation. The shortness of the
decoherence time may present fundamental
difficulties in scaling the quantum
register up to large sizes, which is the basic requirement for
the realization of nontrivial quantum algorithms~\cite{DiVincenzo}.

In this paper, we consider an inductively coupled
charge-qubit model \cite{schon}. Josephson-junction circuits provide two-state
pseudospin systems whose different spin components correspond to
distinct macroscopic variables: either the charges on the
superconducting islands or the phase differences over the Josephson
junctions. Thus, depending on the parameter values for the setup,
one has flux~\cite{pc,Yu}, or charge qubits~\cite{Pashkin,nakamura,vion,martinis,Averin}.
Thus far the largest quantum register, comprising of seven qubits, has been
demonstrated for nuclear magnetic resonance (NMR) in a liquid
solution \cite{Vandersypen}. However, the NMR technique is not
believed to be scalable to much larger registers. In contrast, the
superconducting Josephson-junction circuits are
supposed to provide scalable registers and hence to be better
applicable for large quantum algorithms~\cite{You}.
Furthermore, they allow integration of the control and
measurement electronics. On the other hand, the strong coupling to the
environment through the electrical leads~\cite{Storcz} causes
short decoherence times.

In addition to the quantum register, one needs a quantum gate ``library'', i.e.,
a collection of control parameter sequences
which implements the gate operations on the quantum register.
The quantum gate library must consist of at least a set of universal
elementary gates~\cite{elementary}, which are typically
chosen to be the one-qubit unitary rotations and the CNOT gate.
Some complicated gates may also be included in the library.

The quantum circuit made of these gates resembles
the operational principle of a conventional digital computer.
To minimize the number of gates, the structure
of the quantum circuit can be optimized
using methods similar to those in digital computing~\cite{Aho}.
Minimizing the number of gates is important not only for
fighting decoherence but also for decreasing accumulative errors of classical origin.
If some tailored two-, three- or arbitrary $k$-qubit gates
are included in the gate library, the quantum circuits may be made
much more compact.
The implementation of gates acting on more than two qubits calls for
numerical optimization~\cite{PRL}.
For further discussion on the implementation of non-standard gates
as the building blocks for quantum circuits, see
Refs.~\cite{Burkard,Schuch,Zhang,IJQI}.

We propose an implementation of Shor's algorithm for factoring
moderately large integers - we deal with both algorithmic and
hardware issues in this paper. These are two key aspects of quantum
computation which, however, have traditionally been topics of
disjoint research communities. Hence we aim to provide
a unifying discussion where an expert on quantum algorithms can gain insight
into the realizations using Josephson junctions and experimentalists
working with Josephson devices can choose to read about the quantum
algorithmic aspects. The background material on the construction
of a quantum circuit needed for the evaluation of the modular exponential
function~\cite{Beauregard,Draper} is presented in Appendix A and a derivation of the effective Hamiltonian for
a collection of inductively coupled Josephson qubits is given in Appendix B.

This paper is organized as follows:
The construction of a quantum gate array for Shor's algorithm
is discussed in Sec.~\ref{sec:shor}. In Sec.~\ref{sec:physics},
we consider the Josephson charge-qubit register.
Section~\ref{sec:numerics} presents the numerical methods we have
employed to find the physical implementations of the gates.
Section~\ref{sec:results} discusses in detail how one would realize Shor's
algorithm using Josephson charge qubits to factor the number 21.
Section~\ref{sec:discussion} is devoted to discussion.

\section{\label{sec:shor} Shor's factorization algorithm}

With the help of a quantum computer, one could factor large composite numbers in polynomial
time using Shor's algorithm~\cite{Shor,Ekert,Gruska,Hirvensalo}.
In contrast, no classical polynomial time factorization algorithm is known to date,
although the potential existence of such an algorithm has not been ruled out, either.

\subsection{Quantum circuit}

The strategy for the factoring of a number $N=pq$, both $p$ and $q$ being primes,
using a quantum computer relies on finding the period $r$ of the modular exponential
function $f(x)=a^x \pmod{N}$, where $0<a<N$ is a random number coprime to $N$.
For an even $r$, at least one prime factor of $N$ is given by $\gcd(a^{r/2} \pm1,N)$.
Otherwise, a quantum algorithm must be executed for different values for $a$ until
an even $r$ is found.
\begin{figure}
\begin{picture}(160,85)
\put(-40,0){\includegraphics[width=0.45\textwidth]{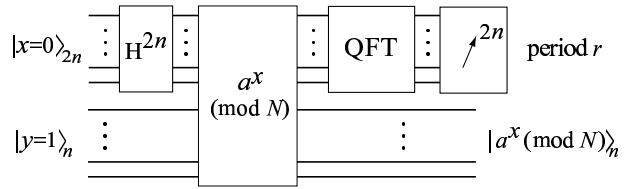}}
\end{picture}
\caption{\label{fg:shor} Quantum circuit for Shor's algorithm. }
\end{figure}

The evaluation of $f(x)$ can be implemented using several
different techniques. To obtain the implementation which involves
the minimal number of qubits, one assumes that the numbers $a$ and $N$
are hardwired in the quantum circuit. However, if a large number
of qubits is available, the design can be easily modified to take
as an input the numerical values of the numbers $a$ and $N$ residing
in separate quantum registers. The hardwired approach combined
with as much classical computing as possible is considerably more
efficient from the experimental point of view.

Figure 1
represents the quantum circuit~\footnote{In the quantum circuit
diagrams, we have indicated the size of a register $|x\rangle_m$
with the subscript $m$.} needed for finding the period $r$. Shor's
algorithm has five stages: (1) Initialization of the quantum
registers. The number $N$ takes $n=\lceil \log_2(N+1) \rceil$ bits
to store into memory, where $\lceil v \rceil$ stands for the
nearest integer equal to or greater than the real number $v$.
To extract the period of
$f(x)$,  we need at least two registers: $2n$ qubits for the register $|x\rangle_{2n}$
to store numbers $x$ and $n$ qubits for the register
$|y\rangle_{n}$ to store the values of $f(x)$. The register
$|x\rangle_{2n}$ is initialized as $|0\rangle_{2n}$, whereas
$|y\rangle_n=|1\rangle_n$. (2) The elegance of a quantum computer
arises from the possibility to utilize arbitrary superpositions. The superposition
state of all integers $0 \leq x \leq 2^{2n}-1$ in the register $|x\rangle_{2n}$ is
generated by applying the Hadamard gate $H$ on each qubit
separately. (3) The execution of the algorithm, the unitary operator
$U_f$, entangles each input value $x$ with the corresponding value
of $f(x)$:

\begin{equation}
U_f\sum_x |x \rangle |1 \rangle=\sum_x |x \rangle |a^x \!\!\!
\pmod{N} \rangle. \label{eq:modexp}
\end{equation}
(4) The quantum Fourier transformation (QFT) is applied
to the register $|x\rangle_{2n}$, which squeezes the probability
amplitudes into peaks due to the periodicity $f(x)=f(x+r)$.
(5) A measurement of the register $|x\rangle_{2n}$ finally
yields an indication of the period $r$.
A repetitive execution of the algorithm reveals the
probability distribution which is peaked at the value $2^{2n}$/$r$
and its integer multiples of output values in the register
$|x\rangle_{2n}$.

Besides the quantum algorithm which is used to find $r$,
a considerable amount of classical precomputing and
postprocessing is required as well. However, all this computing can
be performed in polynomial time.

\subsection{Implementing the modular exponential function}
We are looking for a general scalable algorithm to implement
the required modular exponential function.
The implementation of this part of the algorithm sets limits
for the spatial and temporal requirements of computational
resources, hence it requires a detailed analysis.
The remarkable experimental results \cite{Vandersypen} to
factor the number 15 involve an elegant quantum circuit of
seven qubits and only a few simple quantum gates. The
implementation definitely exploits the special properties of the
number 15, and the fact that the outcome of the function $a^x
\pmod{N}$ can be calculated classically in advance for all input
values $x$ when $N$ is small. For arbitrary $N$, reversible arithmetic
algorithms must be employed~\cite{Beckman,Vedral}.
The classical arithmetic algorithms~\cite{Knuth},
can be implemented reversibly by replacing the irreversible logic gates by their reversible counterparts.
The longhand multiplication algorithm, which we use below, should be
optimal up to very large numbers, requiring only $O(n)$ qubits and
$O(n^3)$ step.

The implementation of the modular exponential
function using a longhand multiplication algorithm and a QFT-based
adder~\cite{Beauregard} provides us with small scratch space
requiring only a total space of $4n+2$ qubits.
The details of the implementation are given in Appendix A.
The conventional approach to longhand multiplication without a QFT-based adder would
require on the order of $5n$ qubits.
The price of the reduced space is the increase in the execution
time, which now is $O(n^4)$, but which can be reduced down to
$O(n^3\log_2 \frac{n}{\epsilon})$, allowing for a certain error
level $\epsilon$. According to Ref.~\cite{Beauregard} one would
achieve an algorithm requiring only $2n+3$ qubits with intermediate measurements. However, we do
not utilize this implementation since the measurements are likely to introduce
decoherence.

\section{\label{sec:physics} Josephson charge-qubit register}

The physical model studied in this paper is the so-called
inductively coupled Cooper pair box array. This model, as well as
other related realizations of quantum computing, has been
analyzed in Ref.~\cite{schon}. The derivation of the
Hamiltonian is outlined in Appendix~\ref{app:deriv} for completeness.
Our approach to quantum gate construction is slightly different from
those found in the literature and it is therefore worthwhile
to consider the physical model in some detail.
\begin{figure}
\includegraphics[width=0.45\textwidth]{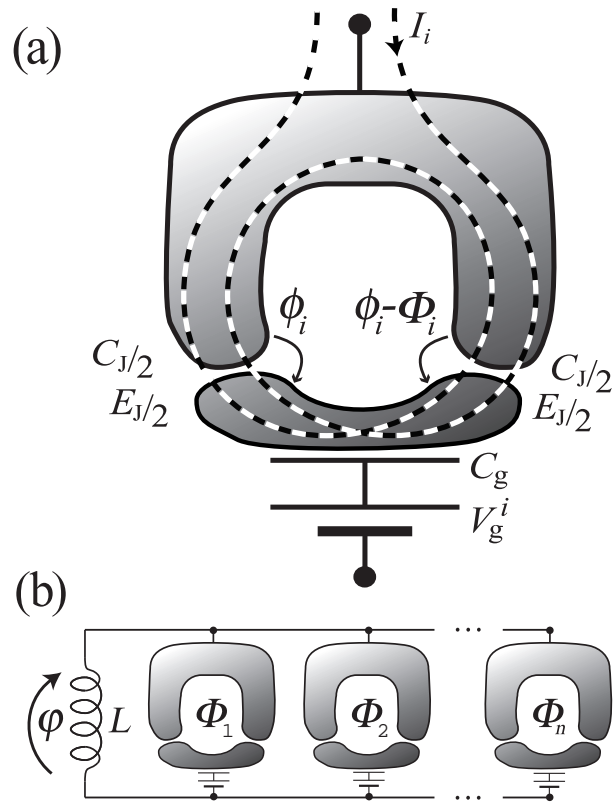}
\caption{\label{fg:kupitit}a) Schematic of a Josephson charge
qubit with the relevant parameters. b) An array of Josephson
charge qubits coupled in parallel with an inductor.}
\end{figure}

A schematic picture of a homogeneous array of qubits
is shown in Fig.~\ref{fg:kupitit}. Each qubit $i$
comprises a superconducting island coupled capacitively to a gate
voltage and a SQUID loop through which Cooper pairs may tunnel.
The gate voltage may be used to tune the effective gate charge
$n_{\rm g}^i$ of the island whereas the external magnetic flux
through the SQUID can be used to control the effective Josephson
energy. Each qubit is characterized by a charging energy $ E_{\rm
C}$ and a tunable Josephson energy $ E_{\rm J}(\Phi_i)$, where
$\Phi_i $ is the flux threading the SQUID. The Hamiltonian for the
$i^{\rm th}$ qubit can be written as
\beq
\label{eq:single}
{\cal H}^i_{\text{single}}=-\frac{1}{2}B^i_z\sigma_z^i-\frac{1}{2}B^i_x\sigma_x^i
\eeq
and the coupling between the $i^{\rm th}$ and  $j^{\rm th}$ qubits as
\beq
\label{eq:couple}
{\cal H}_{\text{coupling}}^{i,j}=-CB^i_xB^j_x\sigma_y^i\otimes
\sigma_y^j \,\,.
\eeq
The qubit state $|0\ket$ (``spin up'')
corresponds to zero extra Cooper pairs residing on the island and the
state $|1\ket$ (``spin down'') corresponds to one extra pair on the
island. Above $B^i_x=E_{\rm J}(\Phi_i)$, $B^i_z=E_{\rm
C}(1-2n_{\rm g}^i)$ and $C=\pi^2L/
\Phi_0^2\left(C_{\rm qb}/C_{\rm J}\right)^2$ denotes
the strength of the coupling between the qubits, whereas $C_{\rm qb}$
is the total capacitance of a qubit in the circuit, $C_{\rm J}$ is
the capacitance of the SQUID, $L$ is the inductance which may in
practice be caused by a large Josephson junction operating in
the linear regime and finally $\Phi_0=h/2e$ is the flux quantum.
The approach taken is to deal with the parameters $B^i_z$ and
$B^i_x$ as dimensionless control parameters. We assume that they
can be set equal to zero which is in principle possible if the
SQUID junctions are identical. We set $C=1$ and choose natural
units such that $\hbar=1$.

The Hamiltonian in Eqs.~(\ref{eq:single}--\ref{eq:couple}) is a
convenient model for studying the construction of quantum
algorithms for a number of reasons. First of all, each
single-qubit Hamiltonian can be set to zero, thereby
eliminating all temporal evolution. Secondly, setting the
effective Josephson coupling to zero eliminates the coupling
between any two qubits. This is achieved by applying half a
flux quantum through the SQUID loops. If the Josephson
energy of any two qubits is nonzero, there will automatically emerge a
coupling between them. This is partly why numerical methods are
necessary for finding the control-parameter sequences. By properly
tuning the gate voltages and fluxes it is possible to compensate
undesired couplings and to perform any temporal evolution in this
model setup.

\begin{figure}
\includegraphics[width=0.45\textwidth]{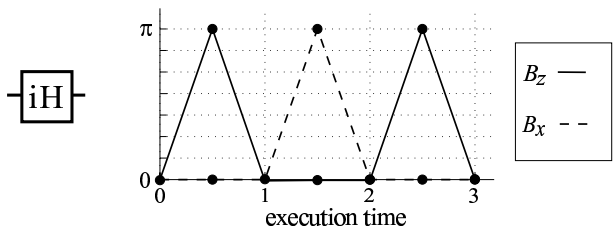}
\caption{\label{fg:Hada} Pulse sequence implementing an
equivalent of the Hadamard gate. Solid line indicates $B_x^i$ while the dashed
line shows $B_z^i$.}
\end{figure}

We note that the generators $i\sigma_x$ and $i\sigma_z$ are
sufficient to construct all the $SU(2)$ matrices through the
Baker-Campbell-Hausdorff formula and thus single-qubit gates need
not be constructed numerically. It is even possible to do this in
a piecewise linear manner avoiding abrupt switching since the only
relevant parameter is the time integral of either $B^i_z$ or
$B^i_x$ if only one of them is nonzero at a time. That is, any
$U\in SU(2)$ acting on the i$^{th}$ qubit can be written as
\beq
\label{eq:SU2}
U=e^{i\sigma_z^i\int_{t_2}^{t_3}B_z^i(t)dt/2}
e^{i\sigma_x^i\int_{t_1}^{t_2}B_x^i(t)dt/2}
e^{i\sigma_z^i\int_{t_0}^{t_1}B_z^i(t)dt/2},
\eeq
 where we assume that from $t_0$ to $t_1$ only $B^i_z$ is
nonzero, from $t_1$ to $t_2$ only $B^i_x$ is nonzero and from
$t_2$ to $t_3$ only $B^i_z$ is again nonzero. For instance, the
gate $iH\in SU(2)$, equivalent to the Hadamard gate $H\in U(2)$ up to a global phase, can be realized as
in Fig.~\ref{fg:Hada} by properly choosing the time-integrals in
Eq.~(\ref{eq:SU2}).
We cannot achieve $U(2^n)$ for $n$ qubits since
the Hamiltonian for the entire quantum register turns out to
be traceless, thus producing only $SU(2^n)$ matrices. However, the global phase factor is not physical
here since it is not observable.

%\subsection{Limitations of the model}

The above Hamiltonian is an idealization and does not take any
decoherence mechanisms into account. To justify this omission, we
have to ensure that a charge-qubit register is
decoherence-free for time scales long enough to execute a
practical quantum algorithm. In addition, we have neglected the
inhomogenity of the SQUIDs. It may be extremely challenging to
fabricate sufficiently uniform junctions. A three-junction design
might alleviate this problem. Whereas for the control of $N$
two-junction SQUIDs one needs at least $N$ independent sources of
flux, the three-junction design would call for $2N$ independent
sources. The extra sources may be used to compensate the
structural nonuniformities. The noise in the control parameters has
also been neglected but it will turn out that the error will grow
linearly with the rms displacement of uncorrelated Gaussian noise.
Correlated noise may only be tolerated if it is very weak. We have
also neglected the issue of quantum measurement altogether in the
above.

A crucial assumption is that $k_{\rm B}T \ln
N_{\rm qp} \ll E_{\rm J} \ll  E_{\rm
C} \ll \Delta_{\rm BCS}$, where $N_{\rm qp}$ is the number of quasiparticle modes.
Typical operation frequencies would be
in the GHz range and the operation temperature could be tens of
mK. For our two-state Hamiltonian to apply, we should actually
insist that, instead of $ E_{\rm J} \ll  E_{\rm C} $, the
requirement $E_{\rm J}(\Phi_i) \ll E_{\rm C} $ holds. It
may appear at first that $B_x^i $ cannot take on values exceeding
$B_z^i$. However, this does not hold since the gate charge also
plays a role; values of $B_z^i$ can be very small if $n_{\rm g}^i$
is tuned close to one half. Since we employ natural units we may
freely rescale the Hamiltonian while rescaling time. This
justifies our choice $C=1$ above. Furthermore, it is always
possible to confine the parameter values within an experimentally
accessible range. For more discussion, see Ref.~\cite{schon}.

\section{\label{sec:numerics} Implementing a Quantum-gate Library}

The evaluation of the time-development operator $U$ is straightforward
once the externally controlled physical parameters for the quantum register
are given. Here we use numerical optimization to solve the inverse problem;
namely, we find the proper sequence for
the control variables which produce the given quantum gate.

\subsection{Unitary time evolution}

The temporal evolution of the Josephson charge-qubit register is described by
a unitary operator
\beq
\label{eq:U}
U_{\gamma(t)}=\mathcal{T}
\exp\left(-i\int_{\gamma(t)} {\cal H}(\gamma(t))dt\right),
\eeq
where $\mathcal{T}$ stands for the time-ordering operator and $ {\cal H}(\gamma(t))$ is the
Hamiltonian for the qubit register. The integration is performed along the path
$\gamma(t)$ which describes the time evolution of the control parameters in the space
spanned by $\{B_x^j(t)\}$ and $\{B_z^j(t)\}$.

Instead of considering paths $\gamma(t)$ with infinitely many degrees of freedom,
we focus on paths parametrized by a finite set of parameters $X_{\gamma}$.
This is accomplished by restricting the path $\gamma(t)$ to polygons in the parameter space.
Since the pulse sequence starts and ends at the origin, it becomes
possible to consistently arrange gates as a sequence.
For an $n$-qubit register, the control-parameter path $\gamma(t)$ is of the vector form
\beq
\gamma(t)=\left[\begin{matrix}B_z^1(t), & \dots & ,B_z^n(t); &
B_x^1(t), & \dots & ,B_x^n(t)\end{matrix}\right]^T,
\eeq
where $B_z^i(t)$ and  $B_z^i(t)$ are piecewise linear functions of time for the chosen parametrization.
Hence, in order to evaluate Eq.~(\ref{eq:U}), one only needs to specify
the $2n$ coordinates for the $\nu$ vertices of the polygon, which we denote collectively as $X_\gamma$.
We let the parameter loop start at the origin, i.e., at the degeneracy point where no time development takes place.
We further set the time spent in traversing each edge of the polygon to be unity.

In our scheme, the execution time for each quantum gate depends linearly on the number $\nu$ of the
vertices in the parameter path. This yields a nontrivial relation between
the execution time of the algorithm and the size of the gates. First note that each $k$-qubit gate represents a matrix in $SU(2^k)$.
To implement the gate, one needs to have enough vertices to parameterize the unitary group $SU(2^k)$, which has $2^{2k}-1$ generators.
In our model, we have $2k$ parameters for each vertex, which implies
$ \label{condition} 2k\nu\geq 2^{2k}-1. $
We have used $\nu=4$ for the two-, and $\nu=11$ for the three-qubit gates.

To evaluate the unitary operator $U_{\gamma(t)}$ we must find a numerical method which is efficient, yet numerically stable.
We divide the path $\gamma(t)$ into tiny intervals that take a time $\Delta t$ to traverse. If $\gamma_i$ collectively denotes
the values of all
the parameters in the midpoint of the $i^{\rm th}$ interval, and $m$ is the number of such intervals, we then find to a good approximation
\beq
U_{X_\gamma}
\approx \exp(-i{\cal H}(\gamma_m)\Delta t) \ldots \exp(-i{\cal H}(\gamma_1)\Delta t) \,.
\label{eq:expapp}
\eeq

We employ the truncated Taylor series expansion
\beq
e^{-i{\cal H} \Delta t}  \approx \sum_{k=0}^l \frac{(-i{\cal H} \Delta t)^k}{k!}
\label{eq:exptay}
\eeq
to evaluate each factor in Eq.~(\ref{eq:expapp}).
We could use the Cayley form
\beq
e^{-i{\cal H} \Delta t}\approx (1-i{\cal H} \Delta t/2)(1+i{\cal H} \Delta t/2)^{-1},
\eeq
or an adaptive Runge-Kutta method to integrate the Schr\"odinger equation as well. It turns out that
the Taylor expansion with $l=3$ is fast and yields enough precision for our purposes.
The precision and unitarity of the approximation are verified by comparing the results with those
obtained with an exact spectral decomposition of ${\cal H}$.

\subsection{Minimization of the error function}
Given an arbitrary matrix $\hat{U}$, our aim is to find a parameter sequence $X_\gamma$ for the Josephson charge-qubit
register that yields a unitary matrix $U_{X_\gamma}=\hat{U}$.
We convert the inverse problem into an optimization task; namely, that of finding the zeroes of the error function
\beq
\label{eq:error} p(X_\gamma)=\|\hat{U}-U_{X_\gamma}\|_{\rm F}.
\eeq
Minimizing $p(X_\gamma)$ over all the possible values of $X_\gamma$
will produce an approximation $U_{X_\gamma}$ for the desired gate $\hat{U}$. Above $\|\cdot\|_{\rm F}$ denotes the
Frobenius trace norm, defined as $\|{A}\|_{\rm F}=\sqrt{\mathrm{Tr}\left({A}^\dagger{A}\right)}$, which
is numerically efficient to compute. Since all the matrix norms are mathematically equivalent,
a small value of $\|{A}\|_{\rm F}$ implies a small value in all other norms as well, see e.g. Ref.~\cite{golub}.

\begin{figure}
\includegraphics[width=0.45\textwidth]{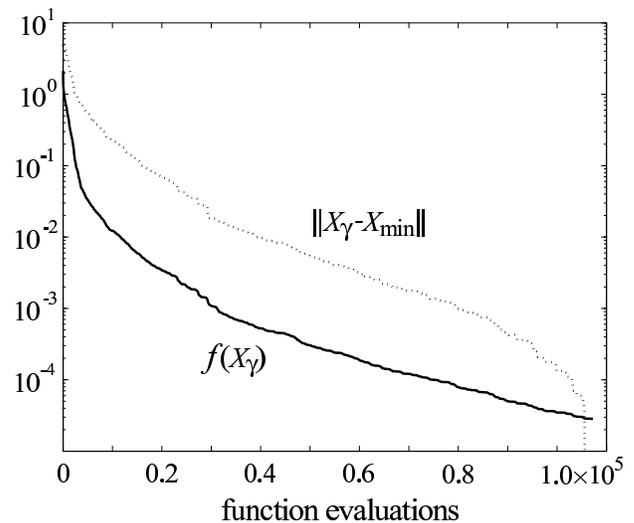}
\caption{\label{fg:convergence}Convergence of the algorithm for the Fredkin gate. The error function values
are indicated by the solid line and the distance of the parameter sequence from the numerical optimum $X_{\rm min}$ by the dotted line.}
\end{figure}

For this minimization problem, the error-function landscape is
rough consisting of many local minima. Consequently, any gradient-based
minimization algorithm will encounter serious problems.
Thus, we have
found the minimum point $X_{\rm min}$ for all the gates presented
in Sec.~\ref{sec:results} using repeated application of a robust
polytope algorithm \cite{IJQI,Reva,poly}.
In the first search, the
initial condition was chosen randomly. At the next stage, the
outcome of the previous search was utilized. In order to
accelerate the evaluation of $U_{\gamma(t)}$ we varied the time
steps $\Delta t$; at an early stage of the optimization a coarse
step was employed while the final results were produced using very
fine steps. Typical convergence of the search  algorithm is
illustrated in Fig.~\ref{fg:convergence}.

The required accuracy for the gate operations is in the range
$10^{-4}$ -- $10^{-5}$ for $p(X_{\gamma})$ for two
reasons: (i) in quantum circuits with a small number of gates, the
total error remains small, and (ii) for
large circuits, quantum-error correction can in principle be
utilized to reduce the accumulated errors~\cite{DiVincenzo}. Our
minimization routine takes on the order of $10^6$ function
evaluations to reach the required accuracy.

\section{\label{sec:results}Example}

To demonstrate the level of complexity for the quantum circuit and the demands on the execution time,
we explicitly present the quantum circuit and some physical implementation for the gates needed
for Shor's algorithm to factor the number $N=21$. We choose $a=11$ and hardwire this into the quantum circuit.

\subsection{Quantum Circuit}
\begin{figure*}
\includegraphics[width=0.95\textwidth]{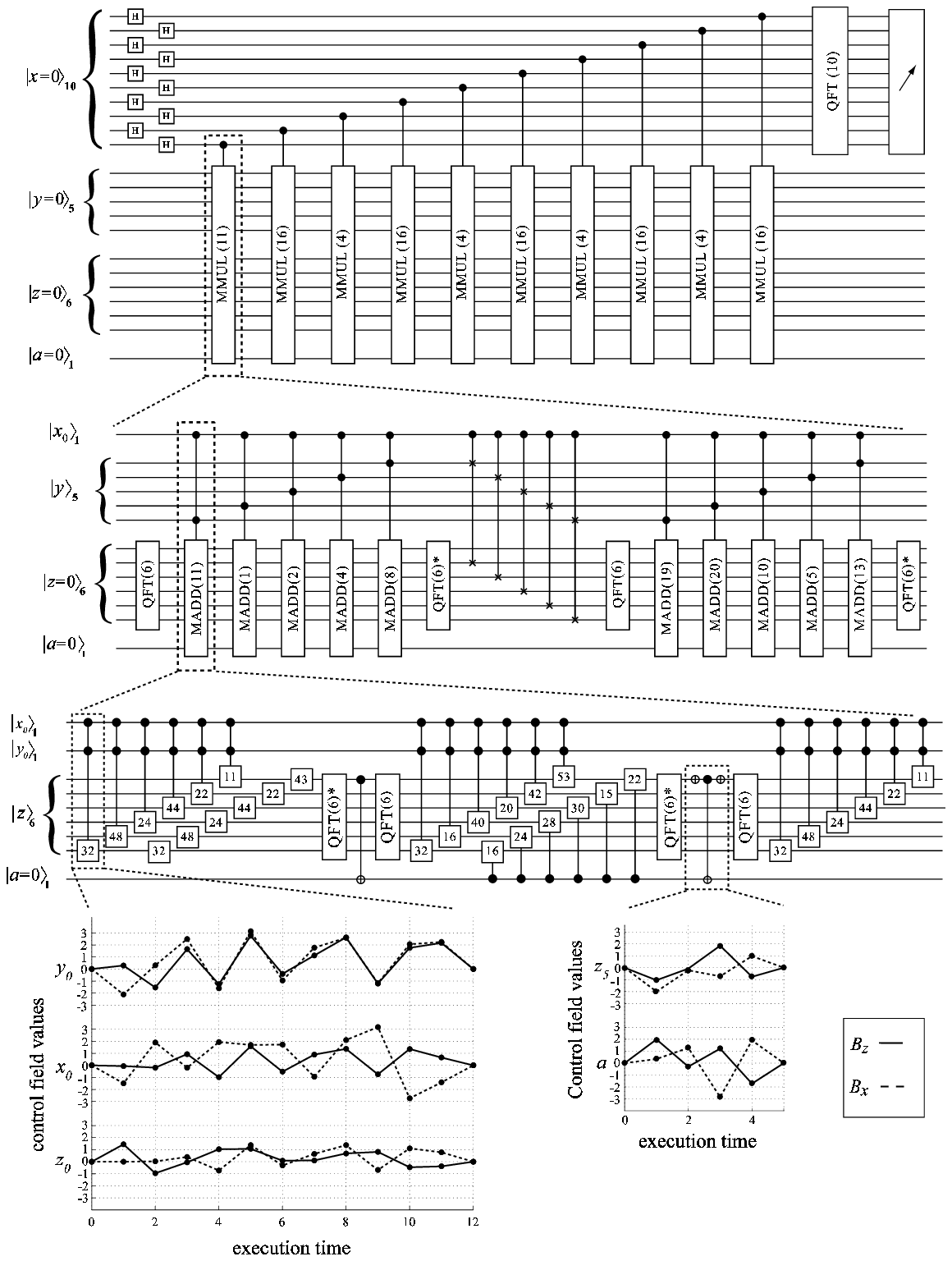}
\caption{\label{virallisuusonilomme} Quantum circuit for Shor's algorithm factoring the number 21 with the parameter value $a=11$. The full circuit is shown topmost and
the decompositions of the modular multiplier and adder blocks are indicated with dashed lines. The gates in the circuit have
their conventional meanings, except that we denote a phase-shift gate by a box with a single number $\phi$ in it meaning that
the phase of the state $|1\ket$ is shifted by $e^{2\pi i \phi/2^n}$ with respect to the state $|0\ket$.
Two examples of
numerically optimized parameter sequences are also shown.}
\end{figure*}

Figure~\ref{virallisuusonilomme} illustrates the structure of the quantum part of the factorization algorithm for the number 21.
Since it takes 5 bits to store the number 21, a 5-qubit register $|y \rangle_5$ and a 10-qubit register $|x \rangle_{10}$ are required.

For scratch space we need a six-qubit register $|z \rangle_6$ and one ancilla qubit $|a\rangle$.
Each thirteen-qubit controlled-MMUL (modular multiplier) gate in the algorithm can be further
decomposed as indicated in Fig.~\ref{virallisuusonilomme}. The controlled-MADD (modular adder) gates can also be
decomposed. The ten-qubit QFT breaks down to 42 two-qubit gates and one three-qubit QFT. Similarly,
the six-qubit QFT can be equivalently implemented as a sequence of 18 two-qubit gates and one three-qubit QFT.
In this manner we can implement the entire algorithm using only one-, two- and three-qubit gates.
The control parameter sequence realizing each of them can then be found using the scheme outlined in Sec.~\ref{sec:numerics}.
Two examples of the pulse sequences are also shown in Fig.~\ref{virallisuusonilomme} (bottom insets).
\begin{figure}
\includegraphics[width=0.45\textwidth]{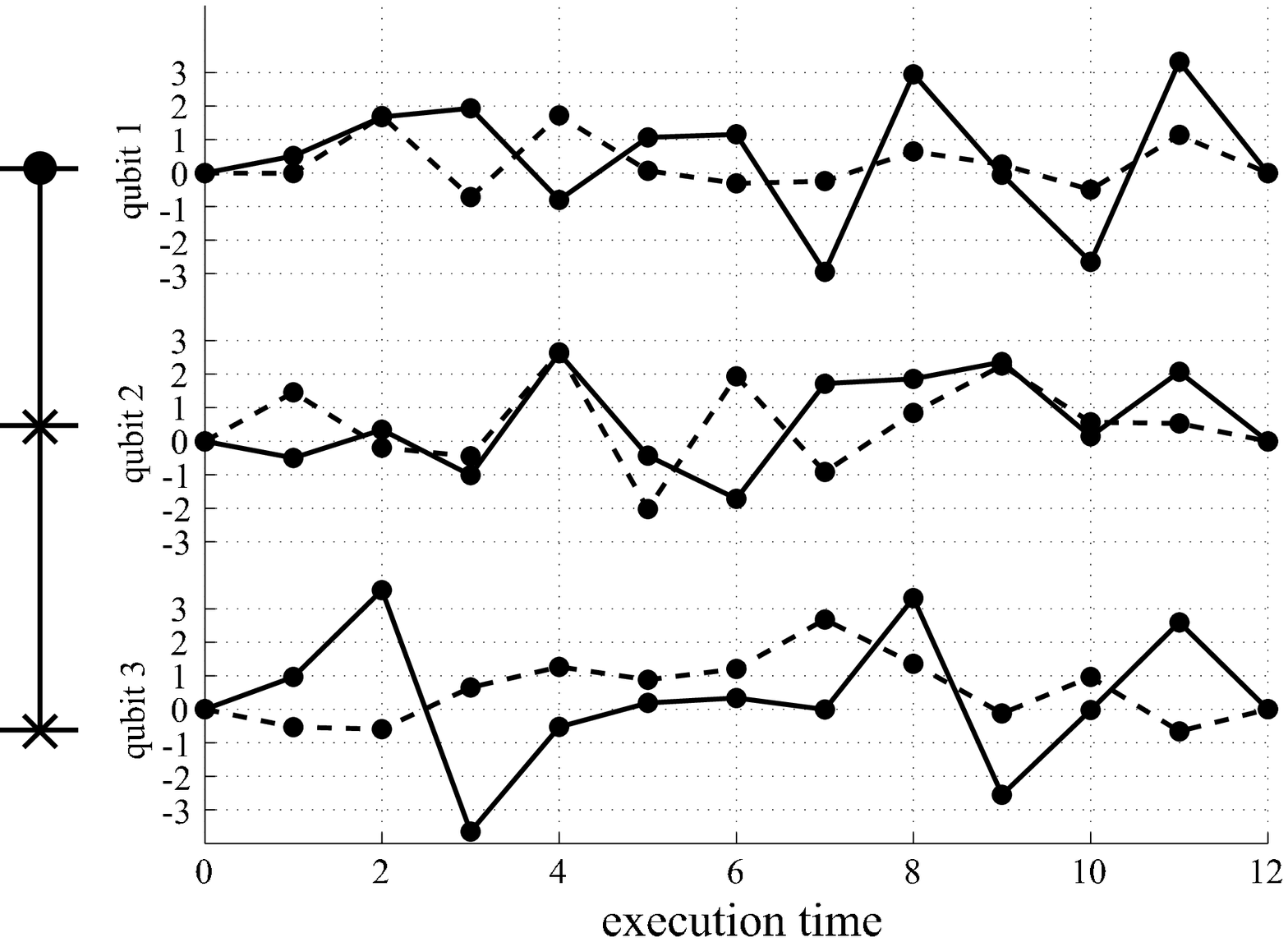}
\caption{\label{fg:Fredkin} Control parameters for the Fredkin
gate. Solid line indicates $B_z^i$ while the dashed line shows
$B_x^i$.}
\end{figure}

\begin{figure}
\includegraphics[width=0.45\textwidth]{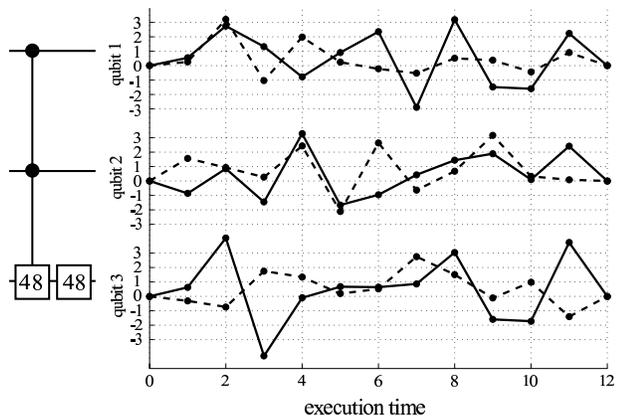}
\caption{\label{fg:Composite} Control parameters for a composite
gate consisting of a controlled$^2$ phase shift and a one-qubit
rotation, see text.  Solid line indicates $B_z^i$ while the dashed
line shows $B_x^i$.}
\end{figure}

% In addition it involves QFT gates for four
%and five qubits. Their implementations are represented in
%Fig.~\ref{fg:four3-5}. Note that the implementation can be
%optimized further by merging the one-qubit phase-shift gates into
%adjacent three-qubit gates.%

%\begin{figure*}
%\includegraphics[width=0.95\textwidth]{15add2b.eps}
%\caption{\label{fg:15add2} Decomposition of a controlled$^2$
%modular adder gate into low-order gates. The gate takes a
%register $|\,\Phi(y)\rangle_5$ as an input, which represents the
%value $y$ in the Fourier basis. The gate is active only if both
%the two control qubits, $| \, x_i \rangle$ and $| \, x_j \rangle$,
%equal $|1 \rangle$. The resulting value $\Phi(y')=\Phi(y+5 \mod
%15)$ if the gate is active, otherwise $\Phi(y')=\Phi(y)$. The
%functionality of the gate requires that the input value $y<N \le
%2^4$.}
%\end{figure*}

%\begin{figure}
%\begin{picture}(160,85)
%\put(-40,0){\includegraphics[width=0.45\textwidth]{add.eps}}
%\end{picture}
%\includegraphics[width=0.45\textwidth]{four3-5.eps}
%\caption{\label{fg:four3-5} Quantum Fourier Transformation for (a) four and (b) five qubits. The $H$ stands for the Hadamard gate.
%The controlled phase shifts are labeled by the numbers $k$ which  correspond to phase shifts by $e^{2\pi k i/2^n}$.
%Note the reversed order of the qubits on the right-hand side.}
%\end{figure}

\subsection{Physical implementation}

The experimental feasibility of the algorithm depends on how
complicated it is compared to the present state of technology.
Following the above construction of the quantum circuit, the full
Shor algorithm to factor 21 requires about 2300 three-qubit gates
and some 5900 two-qubit gates, in total. Also a few one-qubit gates
are needed but alternatively they can all be merged into the multi-qubit gates.
If only two-qubit gates
are available, about 16400 of them are required. If only a minimal
set of elementary gates, say the CNOT gate and one-qubit rotations
are available, the total number of gates is remarkably higher. In
our scheme the execution time of the algorithm is proportional to
the total length of the piecewise linear parameter path which
governs the physical implementation of the gate operations. Each
of the three-qubit gates requires at least a 12-edged polygonal
path $\gamma(t)$ whereas two-qubit gates can be implemented with 5
edges. Consequently, on the order of 57100 edges are required for
the whole algorithm if arbitrary three-qubit gates are available,
whereas $\sim$ 82000 edges would be required for an implementation
with only two-qubit gates.

The ability to find the physical implementation of the gate library for
Shor's algorithm is demonstrated with some further examples.
Figure~\ref{fg:Fredkin} shows how to physically implement the
controlled swap gate. We have taken advantage of tailored
three-qubit implementations: a one-qubit phase-shift gate and a
three-qubit controlled$^2$ phase-shift gate are merged into one
three-qubit gate, see Fig.~\ref{fg:Composite}.

The control parameter sequences presented will yield unitary
operations which approximate the desired gate operations with an
accuracy better than $10^{-4}$ in the error-function values for the three-qubit gates.
For two-qubit gates the error is negligible. Since
the whole factorization circuit consists of some $10^3$ three-qubit gates,
we obtain a total error of $\sim 10^{-1}$. This is sufficient for
the deduction of the essential information from the output. The
robustness of the gates obtained was studied numerically by adding
Gaussian noise to the vertices of the path. The error function was
found to scale linearly with the rms of the variance of the
Gaussian noise: error $\approx$ 6 $\times \langle {\rm noise}
\rangle_{\rm rms}$, which is probably acceptable.

\section{\label{sec:discussion} Discussion}
In this paper we have discussed the implementation of Shor's
factorization algorithm using a Josephson charge-qubit register.
This method is suitable for the first experimental demonstration of
factoring a medium-scale integer $2^4$ -- $2^{20}$.
As an example of this method we have studied the algorithm
for factoring 21.
The only integer smaller than 21 for which Shor's algorithm
is applicable is 15, but this is a special case having only the periods 2 and 4.
For the experimental factoring of 15 one should consider
more direct methods~\cite{Vandersypen} to implement the modular
exponential function.
For a larger integer $N$ other approaches,
e.g. the Sch\"onhage-Strassen multiplication algorithm, will provide a
more efficient quantum circuit. Our approach of numerically
determining the optimized gates can be generalized to other
physical realizations with tunable couplings as well. The only requirement is that the
system allows total control over the control parameters.

We have found that the number of qubits and quantum gates that are
involved in carrying out the algorithm is rather large from the
point of view of current technology. Thus the realization of a
general factorization algorithm for a large integer $N$ will be
challenging. Consequently, the scaling of the chosen algorithm,
both in time and space, will be of prime importance.

The method we propose utilizes three-qubit gates, which
compress the required quantum-gate array, resulting in a shorter
execution time and smaller errors. One should also
consider other implementations of the quantum algorithms that
employ gates acting on a larger number of qubits
to further decrease the number of gates and execution time.
For example, four-qubit gates may be achievable,
but this involves harder numerical optimization.

Finally, let us consider the experimental feasibility of our
scheme. To factor the number 21, we need on the order of $10^4$
edges along the control-parameter path. Assuming that the coherence
time is on the order of 10$^{-6}$~s implies that the upper limit
for the duration of each edge is 10$^{-10}$~s. Since our
dimensionless control parameters in the examples are
on the order of unity, the energy scale in angular frequencies
must be at least on the order of 10$^{10}$ s$^{-1}$. Typical
charging energies for, say, thin-film aluminum structures may be
on the order of $10^{-23}J$ which corresponds to $10^{11}$
s$^{-1}$. The ultimate limiting energy scale is the BCS gap, which
for thin-film aluminum corresponds to an angular frequency of
about $3 \times 10^{11}$ s$^{-1}$. Based on these rough estimates,
we argue that factoring the number 21 on Josephson charge qubits
is, in principle, experimentally accessible.

Constructing a quantum algorithm to decrypt RSA-155 coding
which involves a 512-bit integer $N$ with the scheme that we
have presented would require on the order of $2000$ qubits.
Assuming that the execution time scales as $n^3 \log n$ implies that tens of seconds of decoherence time
is needed. This agrees with the estimates
in Ref.~\cite{Hughes} and poses a huge experimental challenge.
This can be compared to the 8000 MIPS (Million Instructions Per Second)
years of classical computing power which is needed to
decrypt the code using the general numeric
field sieve technique~\cite{Galindo}.
Thus Shor's algorithm does appear impractical for decrypting RSA-155.
However, it provides the only known potentially feasible method to
factor numbers having 1024 or more bits.

We conclude that  it is possible
to demonstrate the implementation of Shor's algorithm on a
Josephson charge-qubit register. Nevertheless, for successful
experimental implementation of large-scale algorithms significant
improvements in coherence times, fabrication and ultrafast control
of qubits is mandatory.

{\it Note added}: After the completion of the revised version of this manuscript, it was
brought to our attention that the implementation of Shor's algorithm
has recently also been considered for a linear nearest-neighbor (LNN)
qubit array model~\cite{Fowler}. In an approach similar to ours, however, the
LLN quantum-circuit model is independent of any specific physical realization.

\acknowledgements
%\section{Acknowledgements}

JJV  thanks the Foundation of Technology (TES, Finland) and the
Emil Aaltonen Foundation for scholarships. MN thanks the Helsinki
University of Technology for a visiting professorship; he  is
grateful for partial support of a Grant-in-Aid from the MEXT and JSPS, Japan
(Project Nos.\ 14540346 and 13135215). The research has been supported
in the Materials Physics Laboratory at HUT by
Academy of Finland through Research Grants in Theoretical Materials
Physics (No. 201710) and in Quantum Computation (No. 206457).
We also thank Robert Joynt, Jani Kivioja, Mikko
M\"ott\"onen, Jukka Pekola, Ville Bergholm, and Olli-Matti Penttinen
for enlightening discussions. We are grateful to CSC -
Scientific Computing Ltd, Finland for parallel computing resources.

\appendix

\section{\label{sec:modexp} Construction of Quantum circuit}

Here we represent the construction of a quantum circuit needed
for an evaluation of the modular exponential function $a^x \pmod{N}$. We assume
the values of $a$ and $N$ to be constant integers coprime to each other.
This approach takes advantage of the well-known fast powers trick, as well as the construction of
a multiplier suggested by Beauregard~\cite{Beauregard}, which in part employs
the adder of Draper~\cite{Draper}.

The modular exponential function can be expressed in terms of modular products:
\begin{equation}
a^x \equiv  \prod_{i=0}^{2n-1} (a^{2^ix_i}  \pmod{N}) \pmod{N},
\label{eq:expdec}
\end{equation}
where we have used the binary expansion
$x=2^0x_0+2^1x_1+2^{n-1}x_{n-1}$, $x_i \in$ \{0,1\}. Note that the
number of factors in Eq.~(\ref{eq:expdec}) grows only linearly for
increasing $n$. The longhand multiplication is based on the relation
\begin{equation}
a^{2^i}x  \equiv \sum_{k=0}^{2n-1} \,(a^{2^i}2^kx_k \pmod{N}) \!\!\!\!
\pmod{N}, \label{eq:muldec}
\end{equation}
which again involves only a linear number of terms.

\begin{figure}
\includegraphics[width=0.45\textwidth]{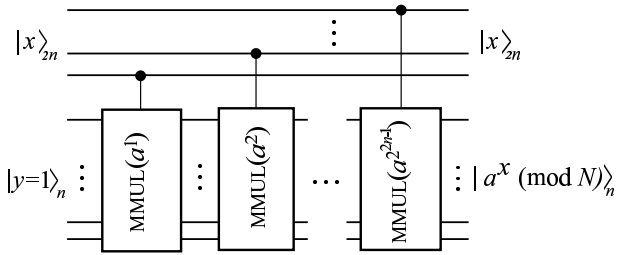}
\caption{\label{fg:mode1} Quantum circuit required for performing
the evaluation of the modular exponential function utilizing the
CMMUL($b$) gates.}
\end{figure}

\begin{figure*}
\includegraphics[width=0.95\textwidth]{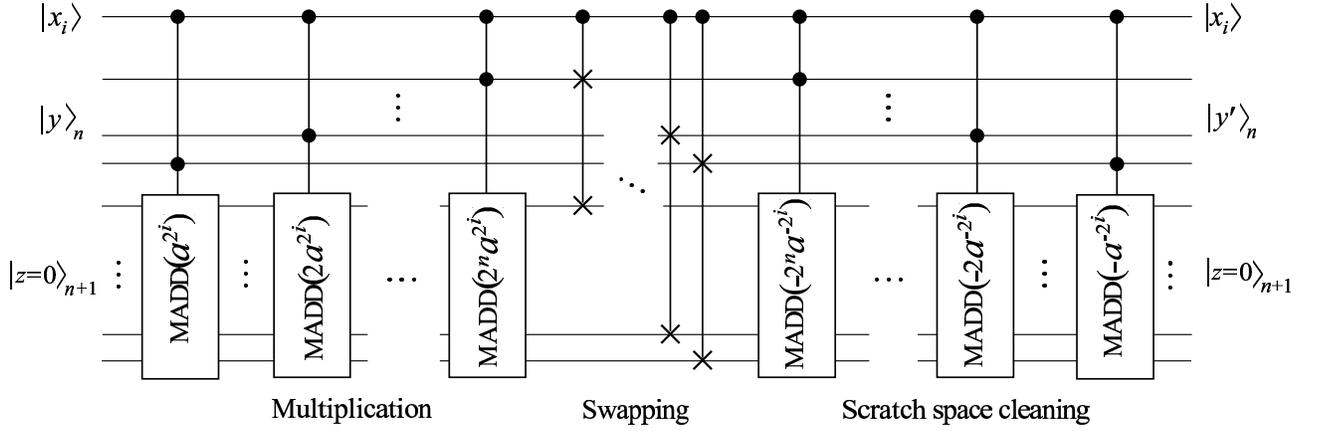}
\caption{\label{fg:mode2} Decomposition of the CMMUL($a^{2^i}$) gate
using C$^2$MADD($b$) and controlled swap gates.
If the controlling qubit $|x_i \rangle$ is active the resulting state is
$y' \equiv  y+a^{2^i} \pmod{N}$, otherwise $y'=y$. Note, that the gate utilizes
an additional ancilla register $|z \rangle_{n}$ to perform the calculation.}
\end{figure*}

Equation~(\ref{eq:expdec}) yields a decomposition of the modular exponential function
into controlled modular multiplication gates (CMMUL($a^{2^i}$)), see Fig.~\ref{fg:mode1}.
According to Eq.~(\ref{eq:muldec}), each of the MMUL($a^{2^i}$) gates can be implemented
with the sequence of the modular adders, see Fig.~\ref{fg:mode2}.
Since this decomposition of CMMUL($a^{2^i}$) requires extra space for
the intermediate results, we are forced to
introduce a scratch space $| z \rangle_{n+1}$ into the setup.
Initially, we set  $|z \rangle_{n+1} = |0 \rangle_{n+1}$. Moreover, we
must reset the extra scratch space after each multiplication.
Euler's totient theorem guarantees that for every $b$ which is coprime to
$N$, a modular inverse $b^{-1} \in \mathbb{N}$ exists.
Furthermore, the extended Euclidean algorithm provides an efficient way to find the numerical
value for $b^{-1}$.

\begin{figure}
\includegraphics[width=0.47\textwidth]{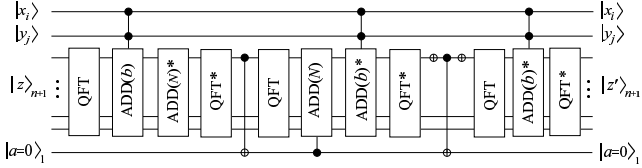}
\caption{\label{fg:mode3} Decomposition of the C$^2$MADD($b$) gate
into elementary gates, QFT gates, and additions in the Fourier
basis (C$^2$ADD). The asterisk stands for a Hermitian conjugate;
it corresponds to a gate for subtraction. The gate takes an input
value $z<N\leq 2^n$ and yields $|z'\rangle_{n+1} = | z+b \pmod{N}
\rangle_{n+1}$ if the control qubits $x_i=1$ and $y_j=1$. Otherwise
$|z'\rangle_{n+1}=|z\rangle_{n+1}$. The ancilla qubit
$|a\rangle$ is one if $z+b>N$ and zero otherwise.}
\end{figure}

\begin{figure}
%\begin{picture}(160,85)
%\put(-40,0){\includegraphics[width=0.45\textwidth]{add.eps}}
%\end{picture}
\includegraphics[width=0.45\textwidth]{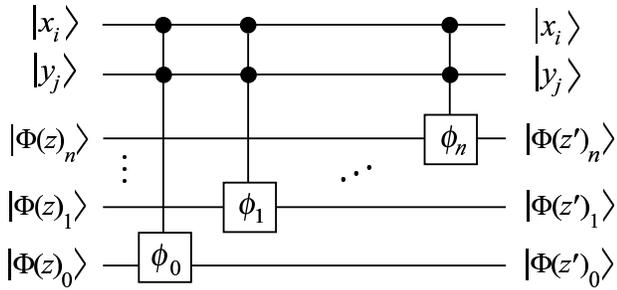}
\caption{\label{fg:add} Quantum circuit for the controlled$^2$
addition of a classical number $b$ into the quantum register
$|z\rangle_{n+1}$ in the Fourier basis. The controlled$^2$ phase-shift
gates serve to yield the phase shift $e^{2 \pi i \phi_k/2^n}$
provided that the control qubits $|x_i\rangle$ and  $|y_j\rangle$
are active.}
\end{figure}

Figure~\ref{fg:mode3} presents decomposition of the C$^2$MADD($b$) gate ($b \in \mathbb{N}$
using adders in the Fourier space. An obvious drawback of this implementation is the need for a number
of  QFT-gates. However, we need to introduce only one ancilla qubit $|a\rangle$.
The decomposition of the gate C$^2$MADD($b$) consists of
controlled$^2$ adders, $(n+1)$-qubit QFTs, one-qubit NOTs, and
CNOTs. The decomposition of a QFT-gate into one- and two-qubit
gates is presented, for instance, in Ref.~\cite{Nielsen}.
Since Fourier space is utilized, the C$^2$ADD($b$)
gates can be implemented \cite{Draper} using controlled$^2$ phase
shifts. The quantum gate sequence for an
adder working in the Fourier space is depicted in
Fig.~\ref{fg:add}. The values of the phase shifts for the gate
C$^2$ADD($b$) are given by $e^{2 \pi i \phi_j /2^n}$, where
$\phi_j=2^jb$.

Finally, we are in the position to perform the unitary
transformation which implements the modular exponential function
using only one-, two- and three-qubit gates. If the three-qubit
gates are not available, further decomposition into one- and
two-qubit gates is needed, see Ref.~\cite{elementary}. For
instance, each three-qubit controlled$^2$U gate decomposes into
five two-qubit gates and each Fredkin gate takes seven two-qubit
gates to implement.


%****************************************************************

\section{\label{app:deriv}Derivation of the Hamiltonian}

\subsection{The Lagrangian}

Consider a homogenous array of mesoscopic superconducting islands
as an idealized model of a quantum register, see
Fig.~\ref{fg:kupitit}. The basis states of the qubit correspond to
either zero or one extra Cooper pair residing on the
superconducting island,  denoted by $|0\ket$ and $|1\ket$,
respectively.  Each of the islands, or Cooper-pair boxes, is
capacitively coupled to a gate voltage, $V_{\rm g}^i$. In
addition, they are coupled to a superconducting lead through a
mesoscopic SQUID with identical junctions, each having the same
Josephson energy $E_{\rm J}/2$ and capacitance $C_{\rm J}/2$. All
these qubits are then coupled in parallel with an inductor, $L$.
The lowest relevant energy scale is set by the thermal energy
$k_{\rm B}T$ and the highest scale by the BCS gap $\Delta_{\rm
BCS}$.

We assume that the gate voltage $V_{\rm g}^i$ and the
time-dependent flux $\Phi_i$ through each SQUID can be controlled
externally. The flux $\Phi_i$ may be controlled with an adjustable
current $I_i$ through an external coil, see the dotted line in
Fig.~\ref{fg:kupitit}a. In this setup, the Cooper pairs can tunnel
coherently to a superconducting electrode. We denote the time-integral of voltage, or
difference in flux units, over the left junction of the
$i^{\rm th}$ SQUID by $\phi_i$ and the flux through the
inductor by $\varphi$. The phase difference in flux units over the rightmost
junction is $\phi_i-\Phi_i$. We take the positive direction for
flux to be directed outward normal to the page.

We adopt $\phi_i$ and $\varphi$ as the dynamical variables,
whereas $\Phi_i$ and $V_{\rm g}^i$ are external adjustable
parameters. With the help of elementary circuit
analysis~\cite{Devoret}, we obtain the Lagrangian for the qubit
register
\begin{align}
\mathcal{L}& =\frac{1}{2}\sum_{i=1}^n\left[\frac{C_{\rm
J}}{2}\dot{\phi}_i^2
+\frac{C_{\rm J}}{2}(\dot{\phi}_i-\dot{\Phi}_i)^2+C_{\rm g}(\dot{\phi}_i+\dot{\varphi}-V_{\rm g}^i)^2 \right] \nonumber \\
&-\frac{\varphi^2}{2L}+\frac{1}{2}\sum_{i=1}^n\left[E_{\rm
J}\cos\left(\frac{2e}{\hbar}\phi_i\right) + E_{\rm
J}\cos\left(\frac{2e}{\hbar}(\phi_i-\Phi_i)\right)\right].
\label{eq:Lag}
\end{align}
We now perform the following changes of variables \beq \phi_i \to
\phi_i+\frac{\Phi_i}{2}-\frac{C_{\rm g}}{C_{\rm J}+C_{\rm
g}}\varphi \eeq
%This will serve two purposes. It will diagonalize
%the Lagrangian (as well as the corresponding Hamiltonian) and it
%will also allow us to rewrite the potential energy contribution in
%Eq.~(\ref{eq:Lag}) in a more transparent form. It is convenient to
%carry out these transformations before proceeding to the
%Hamiltonian formalism since in this way we need not consider any
%commutation relations or modifications to the derivative terms due
%to the interaction picture.
%The change of variables
which yields
\begin{align}
\mathcal{L}& =\frac{1}{2}\sum_{i=1}^n\left[(C_{\rm J}+C_{\rm
g})\dot{\phi}_i^2
-2C_{\rm g}\left(V_{\rm g}^i-\frac{\dot{\Phi}_i}{2}\right) \dot{\phi}_i \right. \nonumber \\
& \left.+E_{\rm
J}\cos\left(\pi\frac{\Phi_i}{\Phi_0}\right)\cos\left(\frac{2e}{\hbar}\phi_i
-\frac{2\pi C_{\rm qb}}{\Phi_{0} C_{\rm J}}\varphi\right)\right] \nonumber \\
&+\frac{1}{2}NC_{\rm qb}\dot{\varphi}^2-\sum_{i=1}^{n}C_{\rm
qb}(V_{\rm g}^i-\frac{\dot{\Phi}_i}{2})\dot{\varphi}
-\frac{\varphi^2}{2L} \\
&+{\rm const}. \label{eq:Lag2}
\end{align}
Above, $\Phi_0=h/2e$ is the flux quantum and $C_{\rm
qb}=C_{\rm J}C_{\rm g}/(C_{\rm J}+C_{\rm g})$ is the
qubit capacitance in the $LC$-circuit. Note that the effective
Josephson energy of each SQUID can now be tuned. We denote this
tunable energy parameter in Eq.~(\ref{eq:Lag2}) as \beq E_{\rm
J}(\Phi_i)=E_{\rm J}\cos\left(\pi\frac{\Phi_i}{\Phi_0}\right).
\eeq The canonical momenta are given by $Q =\partial\mathcal{L}/\partial \dot{\varphi}$
and $q_i = \partial\mathcal{L}/\partial \dot{\phi}_i$.
We interpret $Q$ as the charge on the collective capacitor
formed by the whole qubit register, whereas $q_i$ is the charge on
the $i^{\rm th}$ island. Note that the charge $q_i$ is related to
the number $n_i$ of Cooper pairs on the island through
$q_i=-2en_i$.

\subsection{The Hamiltonian}
We are now in the position to write down the Hamiltonian for the
quantum register. We will also immediately replace the canonical
variables by operators in order to quantize the register.
Moreover, we will employ the number of excess Cooper pairs $n_i$ on the
island and the superconducting phase difference instead of the
usual quantum-mechanical conjugates. We will also change to the
more common phase difference $\theta_i$ related to $\phi_i$
through $\theta_i=\frac{2e}{\hbar}\phi_i$. Hence the relevant
commutation relations are $ [\theta_i,n_i]=-i $ and
$ [\varphi,Q]=i\hbar$. All the other commutators vanish.
Using the Legendre transformation \beq
\mathcal{H}=Q\dot{\varphi}+\sum_{i=1}^{n}q_i
\dot{\phi}_i-\mathcal{L} \eeq we obtain
\begin{align}\label{eq:ham1}
\mathcal{H} &=\sum_{i=1}^{n}\left[\frac{2e^2(n_i-n_{\rm
g}^i)^2}{C_{\rm J}+C_{\rm g}} -E_{\rm J}(\Phi_i)\cos\left(\theta_i
-\frac{2\pi C_{\rm qb}}{\Phi_{0} C_{\rm J}}\varphi\right) \right] \nonumber \\
&+\frac{ (Q+Q_{\rm g} )^2}{2NC_{\rm qb}}+\frac{\varphi^2}{2L}.
\end{align}
We have denoted the effective gate charge by \beq n_{\rm
g}^i=\frac{C_{\rm g}}{2e}\left(V_{\rm
g}^i-\frac{\dot{\Phi}_i}{2}\right) \eeq and
\beq Q_{\rm
g}=\sum_{i=1}^{n} C_{\rm qp}\left(V_{\rm
g}^i-\frac{\dot{\Phi}_i}{2}\right). \eeq
In addition to the usual voltage contribution, the time
dependence of the flux also plays a role. In practice, the rates
of change of the flux are negligible in comparison to the voltages
and this term may safely be dropped.

The Hamiltonian in Eq.~(\ref{eq:ham1}) describes the register of
qubits ($n_i,\phi_i$) coupled to a quantum-mechanical
$LC$-resonator, i.e., a harmonic oscillator  ($Q,\varphi$). We
will now assume that the rms fluctuations of $\varphi$ are small
compared to the flux quantum $\Phi_0$ and also that the harmonic
oscillator has a sufficiently high frequency, such that it
stays in the ground state. The first assumption implies that \beq
\cos\left(\theta_i-\frac{2\pi C_{\rm qb}}{\Phi_{0} C_{\rm
J}}\varphi\right)\approx \cos\theta_i+\frac{2\pi C_{\rm
qb}}{\Phi_{0} C_{\rm J}}\varphi \sin \theta_i \,\,\,. \eeq The
second assumption will cause an effective coupling between the
qubits. Namely, the Hamiltonian may now be rewritten in the more
suggestive form
\begin{align}\label{eq:ham2}
\mathcal{H} & \approx \sum_{i=1}^{n}\left[\frac{2e^2(n_i-n_{\rm
g}^i)^2}{C_{\rm J}+C_{\rm g}} -E_{\rm J}(\Phi_i)\cos \theta_i
\right] \nonumber \\
&+\frac{ (Q+Q_{\rm g} )^2}{2NC_{\rm qb}}+\frac{\left(
\varphi-\hat{\varphi} \right)^2}{2L}-\frac{\hat{\varphi}^2}{2L},
\end{align}
where the operator $\hat{\varphi}$ is given by \beq
\hat{\varphi}=\frac{2\pi L C_{\rm qb}}{\Phi_{0} C_{\rm
J}}\sum_{i=1}^{n}E_{\rm J}(\Phi_i)\sin\theta_i \,\,\,. \eeq We now
see from Eq.~(\ref{eq:ham2}) that in the high-frequency limit the
harmonic oscillator is effectively decoupled from the qubit
register. The effect of the qubit register is thus to redefine the
minimum of the potential energy for the oscillator. This does not
affect the spectrum of the oscillator, since it will adiabatically
follow its ground state in the low-temperature limit. We may
therefore trace over the degrees of freedom of the harmonic
oscillator and the harmonic-oscillator energy will merely yield a
zero-point energy contribution, $\hbar\omega_{LC}/2$ . The
effective Hamiltonian describing the dynamics of the coupled qubit
register alone is thus
\begin{align}\label{eq:ham3}
\mathcal{H} & \approx \sum_{i=1}^{n}\left[\frac{2e^2(n_i-n_{\rm g}^i)^2}{C_{\rm J}+C_{\rm g}} -E_{\rm J}(\Phi_i)\cos \theta_i \right] \nonumber \\
&-\frac{2\pi^2 L C_{\rm qb}^2}{\Phi_{0}^2 C_{\rm
J}^2}\left(\sum_{i=1}^{n}E_{\rm J}(\Phi_i)\sin\theta_i\right)^2.
\end{align}
This result is in agreement with the one presented in
Ref.~\cite{schon}. We conclude that the $LC$-oscillator has
created a virtual coupling between the qubits.

For the purposes of quantum computing, it is convenient to
truncate the Hilbert space such that each Cooper-pair box will
have only two basis states. In the limit of a high charging energy
$E_{\rm C}=2e^2/(C_{\rm g}+C_{\rm J})$ relative to the Josephson
energy $E_{\rm J}$, we may argue that in the region $0 \leq
n^i_{\rm g} \leq 1$ only the states with $n_i=0,1$ can be
occupied. We use the vector representation for these states, in
which
$|0\ket_i= \left(\begin{matrix}1 & 0
\end{matrix}\right)^{\rm T}_i $ and $ |1\ket_i=\left(\begin{matrix}0
& 1 \end{matrix}\right)^{\rm T}_i$.

The basis states of the Hilbert
space are orthogonal $\bra n |e^{\pm i\theta}|m
\ket=\delta_{n,m\mp 1}$. Hence, in this two-state approximation,
$\cos\theta_i=\frac{1}{2}\sigma_x^i$ and
$\sin\theta_i=\frac{1}{2}\sigma_y^i$, where, e.g.,
$\sigma_x^i=\underbrace{I \otimes \ldots \otimes I}_{i-1 \;{\rm
times}}\otimes\sigma_x\otimes \underbrace{I \otimes I \ldots
\otimes I}_{N-i \;{\rm times}}$. Finally, omitting the constant
terms, we obtain the Hamiltonian in the Pauli-matrix
representation
\begin{align}\label{eq:ham4}
\mathcal{H}_{\rm qb} &=\sum_{i=1}^{n}
\left[-\frac{E_{\rm C}}{2}(1-2n_{\rm g}^i)\sigma_z^i-\frac{E_{\rm J}(\Phi_i)}{2}\sigma_x^i  \right] \nonumber \\
&-\frac{\pi^2 L}{\Phi_0^2}\left(\frac{C_{\rm qb}}{C_{\rm
J}}\right)^2 \sum_{i=1}^{n}\sum_{j=i+1}^{n}E_{\rm J}(\Phi_i)E_{\rm
J}(\Phi_j)\sigma_y^i\otimes\sigma_y^j\,\,,
\end{align}
which results in Eqs. (\ref{eq:single}) and (\ref{eq:couple}) of the main text.

\end{document}